\documentclass[aps,prb,superscriptaddress,twocolumn,showpacs,preprintnumbers,amsmath,amssymb]{revtex4}

\usepackage{graphicx}
\usepackage{epstopdf}
\usepackage{longtable}
\usepackage{bm}
\usepackage{multirow}
\usepackage{hyperref}
\usepackage{array}
\usepackage{booktabs}
\usepackage{amssymb}
\usepackage{amsmath}

\begin{document}

\title{Lattice dynamics of $A$Sb$_2$O$_6$ ($A=$ Cu, Co) with trirutile structure}

\author{D. T. Maimone}
\affiliation{``Gleb Wataghin'' Institute of Physics, University of Campinas - UNICAMP, Campinas, S\~ao Paulo 13083-859, Brazil}

\author{A. B. Christian}
\affiliation{Department of Physics, Montana State University, Bozeman, Montana 59717, USA}

\author{J. J. Neumeier}
\affiliation{Department of Physics, Montana State University, Bozeman, Montana 59717, USA}

\author{E. Granado}
\affiliation{``Gleb Wataghin'' Institute of Physics, University of Campinas - UNICAMP, Campinas, S\~ao Paulo 13083-859, Brazil}

\begin{abstract}

Raman spectroscopy experiments on single crystals of CuSb$_2$O$_6$ and CoSb$_2$O$_6$ quasi-one-dimensional antiferromagnets with trirutile crystal structure were performed, with a focus on the first material. The observed Raman-active phonon modes and previously reported infrared-active modes were identified with the aid of {\it ab}-initio lattice dynamics calculations. The structural transition between monoclinic $\beta$-CuSb$_2$O$_6$ and tetragonal $\alpha$-CuSb$_2$O$_6$ phases at $T_s=400$ K is manifested in our spectra by a ``repulsion'' of two accidentally quasi-degenerate symmetric modes below $T_s$, caused by a phonon mixing effect that is only operative in the monoclinic $\beta$-CuSb$_2$O$_6$ phase due to symmetry restrictions. Also, two specific phonons, associated with CuO$_6$ octahedra rotation and with a Jahn-Teller elongation mode, soften and broaden appreciably as $T \rightarrow T_s$. A crossover from a displacive to an order-disorder transition at $T_s$ is inferred.

\end{abstract}

\pacs{78.30.-j, 63.20.D-, 64.60.Cn, 63.20.kg}

\maketitle

\section{Introduction}

\begin{figure*}
	\includegraphics[width=1.0 \textwidth]{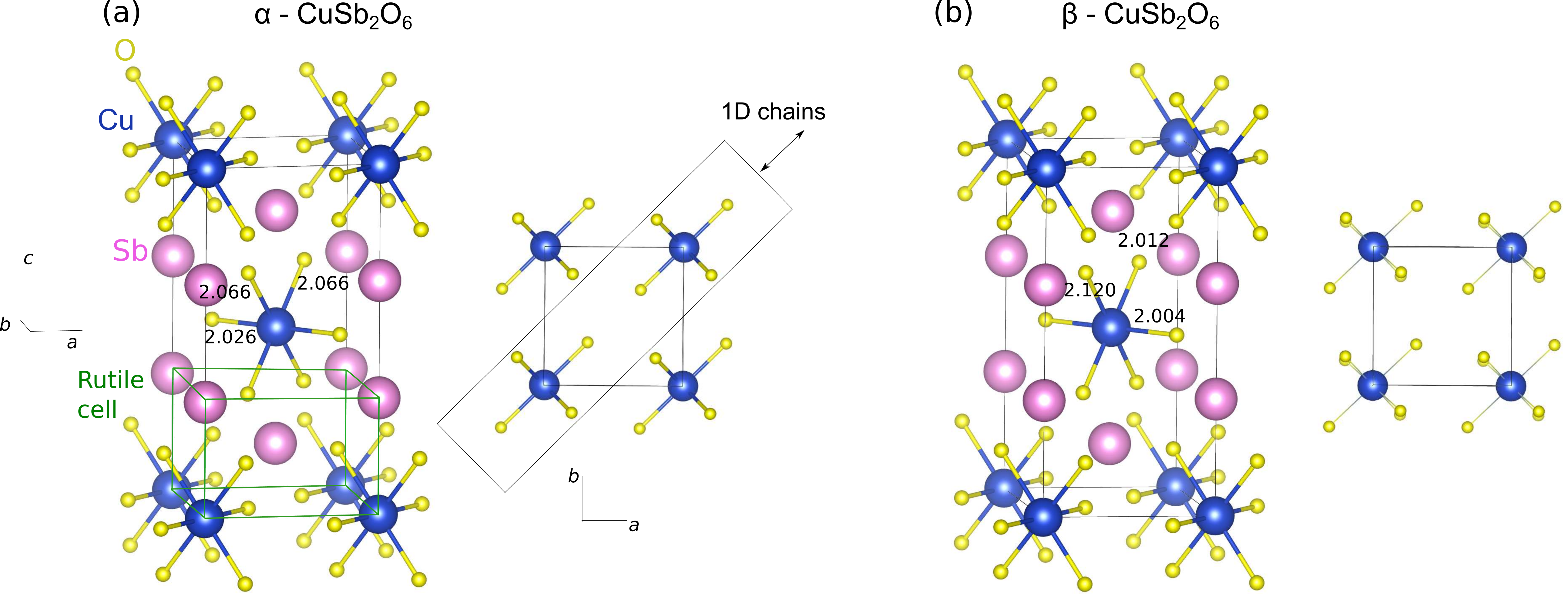}
	\caption{\label{Structure} Crystal structure of (a) tetragonal $\alpha$-CuSb$_2$O$_6$ at $T=400$ K (Ref. \onlinecite{Giere}) and (b) monoclinic $\beta$-CuSb$_2$O$_6$ at room temperature (Ref. \onlinecite{Nakua}). Three-dimensional trirutile crystal structures and {\it ab}-plane projected CuO$_6$ layers with the Cu$^{2+}$ ions at $z=0$ are indicated. The relevant Cu-O distances are given in Angstroms. The unit cell of the parent rutile structure and the exchange direction for the quasi-one-dimensional (1D) spin chains at $z=0$ are indicated in (a).}
\end{figure*}

Low dimensional magnetic systems have attracted continuous interest due to their enhanced quantum fluctuation effects. Nearly ideal one-dimensional (1D) or 2D magnetic behavior may be experimentally attained in crystalline materials that show dominant exchange interactions along a specific direction or within a plane. Examples of quasi-1D magnetism are found in materials that crystallize in the trirutile structure with chemical formula $AB_2$O$_6$ ($A$=magnetic, $B$=non-magnetic cations). This structure is derived from the TiO$_2$ rutile structure by allowing distinct $A$ and $B$ metallic sites and triplicating the unit cell along the $c$ direction (see Fig. \ref{Structure}). Perhaps the most studied material of this class is the spin-$1/2$ quantum spin chain CuSb$_2$O$_6$.\cite{Donaldson,Nakua,Nakua2,Yamagushi,Giere,Kato,Koo,Kato2,Prokofiev,Heinrich,
Torgashev,Gibson,Wheeler,Kasinathan,Rebello,Christian,Prasai,Herak} 
In this material, the Cu$^{2+}$ spins show a 3D magnetic order below $T_N^{3D} \sim 8.5$ K. The magnetic susceptibility curve has a broad maximum at about 60 K, being well modelled by a 1D Bonner and Fisher model\cite{Bonner} with an Heisenberg exchange coupling strength $J \sim -100$ K between $S=1/2$ spins.\cite{Nakua,Yamagushi,Kato,Kato2,Prokofiev,Heinrich,Torgashev,Gibson,Rebello}
This value of $J$ has been associated with  antiferromagnetic Cu-O-O-Cu superexchange interactions along the [1 1 0] and [1 -1 0] directions for Cu ions at $z=0$ and $z=1/2$, respectively [see Fig. \ref{Structure}(a)], while the interchain exchange interactions are estimated to be at least two orders of magnitude weaker.\cite{Nakua,Kasinathan} Another representative material of this class is CoSb$_2$O$_6$ with $T_N^{3D} \sim 13$ K and a maximum at $\sim 35$ K in the magnetic susceptibility curve,\cite{Christian,Reimers,Kato,Christian2} also indicating low-dimensional magnetic correlations above $T_N^{3D}$. While an early report stated that these data are consistent with the anisotropic square planar Ising model, more recent studies rather indicate the presence of 1D Ising chains of Co$^{2+}$ $S=3/2$ spins for CoSb$_2$O$_6$ (Refs. \onlinecite{Christian,Christian2}), with the intrachain exchange constant $J=-10.6$ K being substantially smaller than in CuSb$_2$O$_6$.

The overall magnetic behavior of materials are often determined by details of their crystal structure. CuSb$_2$O$_6$ shows a structural phase transition at $T_s \sim 400$ K \cite{Donaldson,Giere,Heinrich} between tetragonal [$\alpha$-CuSb$_2$O$_6$, space group $P4_2/mnm$, see Fig. \ref{Structure}(a)]\cite{Giere} and monoclinic [$\beta$-CuSb$_2$O$_6$, space group $P2_1/n$, see Fig. \ref{Structure}(b)]\cite{Nakua} structural variants. The high-temperature $\alpha$-CuSb$_2$O$_6$ phase shows slightly compressed CuO$_6$ octahedra along the 1D chain direction,\cite{Giere} while the $\beta$-CuSb$_2$O$_6$ phase shows elongated octahedra, with the elongation axis being nearly perpendicular to the 1D magnetic chains (see Fig. \ref{Structure}). Other similar compounds with Jahn-Teller active ions, such as CrTa$_2$O$_6$ (Ref. \onlinecite{Saes}) and CuTa$_2$O$_6$ (Ref. \onlinecite{Golubev}) also exhibit analogous monoclinic distortions. Besides the Jahn-Teller distortion, the monoclinic phase of CuSb$_2$O$_6$ exhibits a subtle rotation of the CuO$_6$ octahedra (see right pannel of Fig. \ref{Structure}(b)], which contributes to relax the lattice strain. The structural transition at $T_s$ is reported to be of second-order.\cite{Donaldson,Giere,Heinrich} CoSb$_2$O$_6$, on the other hand, crystallizes in a tetragonal lattice with no structural phase transition between 4 and 300 K.\cite{Donaldson,Reimers}

Vibrational spectroscopies including Raman scattering and infrared absorption are powerful tools to investigate structural, electronic and magnetic phase transitions, since lattice vibrations may be sensitive to slight electronic and structural modifications, also sensing magnetic correlations through the spin-phonon coupling mechanism.\cite{Granado} For CuSb$_2$O$_6$ in particular, the 3D magnetic transition at $T_N^{3D}$ has been suggested to be associated with a spin-Peierls transition occurring as a result of three-dimensional phonons coupling with Jordan-Wigner-transformed fermions,\cite{Rebello} and thermal conductivity reveals strong phonon-spin resonant scattering,\cite{Prasai} reinforcing the need for detailed investigations of the vibrational properties of this and other materials with trirutile structure. An important step towards this direction is to achieve a reliable identification of the corresponding normal modes of vibration. This is a challenging task, considering that this structure, with 18 atoms per unit cell, allows for 54 vibrational degrees of freedom. This issue was originally addressed by Husson, Repelin and Brusset\cite{Husson} and Haeuseler,\cite{Haeuseler} however the mode frequencies and mechanical representations are highly dependent on the model and on the chosen values of force fields, leading to considerable ambiguity in the mode assignment. This ambiguity was reduced but not fully eliminated by a more recent group theoretical treatment that related the normal modes of the trirutile structure with those of the simpler rutile one.\cite{Torgashev}

Despite the challenges in assigning the observed vibrational modes, Raman and infrared spectroscopies have been continuously employed to investigate trirutile-like materials.\cite{White,Franck,Husson,Haeuseler,Giere,Torgashev,Bonilla,Golubev} Temperature-dependent Raman spectra of CuSb$_2$O$_6$ showed an evolution of a peak at 645 cm$^{-1}$ at $T=433$ K ($>T_s$) into two peaks, at 639 and 672 cm$^{-1}$, at $T=293$ K ($<T_s$),\cite{Giere} which, supported by the mode assignment of Ref. \onlinecite{Husson}, was ascribed to a degenerate $E_g$ mode of the tetragonal phase being splitted in two separate modes in $\beta$-CuSb$_2$O$_6$. Remarkably, such mode splitting at $T_s$ was not observed in the related material CuTa$_2$O$_6$.\cite{Golubev}
An infrared absorption study showed an
agreement between the number of predicted and observed phonon modes for $\beta$-CuSb$_2$O$_6$, also reporting noticeable anomalies for some phonons around $T=50$ K that are presumably associated with the presence of magnetic correlations.\cite{Torgashev}

Structural phase transitions have been widely investigated by vibrational spectroscopies, and may be classified either as displacive type, with at least one associated soft mode with frequency $\omega \rightarrow 0$ as $T \rightarrow T_s$ (Ref. \onlinecite{Cochran}), or order-disorder type, with large broadening of some phonon peaks.\cite{GranadoJT} Cases with a crossover between displacive and order-disorder types were also reported.\cite{Bruce,Muller} Structural transitions associated with Jahn-Teller active ions are expected to be of order-disorder character if the Jahn-Teller energy level splitting is much larger than the stiffness parameter that orients the distortion from site to site,\cite{Millis} leading to local, non-cooperative Jahn-Teller distortions above $T_s$ such as in LaMnO$_3$.\cite{GranadoJT,EXAFSJT} On the other hand, when the Jahn-Teller level splitting is relatively small, the structural phase transition may be of displacive type with the presence of a soft Jahn-Teller mode.\cite{Srinivasan} For the specific case of CuSb$_2$O$_6$, the relatively small Jahn-Teller distortion of the CuO$_6$ octahedra [see Fig. \ref{Structure}(b)] is presumably associated with a small separation between the $3d_{3z^2-r^2}$ and $3d_{x^2-y^2}$ orbital levels,\cite{Kasinathan} and the order-disorder or displacive character or the structural phase transition remains to be determined. 

In this work, we report temperature-dependent Raman scattering data on single crystalline samples of CuSb$_2$O$_6$ and CoSb$_2$O$_6$, complemented by {\it ab}-initio lattice dynamics calculations. The Raman-active modes observed in this work and the infrared-active modes reported in Ref. \onlinecite{Torgashev} are identified and the corresponding mechanical representations are obtained. The Jahn-Teller transition of CuSb$_2$O$_6$ is also investigated in detail. Two Raman-active modes,  associated with CuO$_6$ octahedra rotation and with a Jahn-Teller elongation mode, soften appreciably as $T \rightarrow T_s$ from below, indicating an incipient behavior characteristic of a displacive phase transition. However, this displacive mechanism is truncated by a strong phonon damping, signaling a crossover to an order-disorder transition at $T_s$. This transition is also manifested in our data by a remarkably large phonon mixing effect that occurs in the spectral region $620 < \omega < 680$ cm$^{-1}$ in the monoclinic $\beta$-CuSb$_2$O$_6$ phase. The phonon assignment and investigation of the structural transition of CuSb$_2$O$_6$ achieved in this work pave the way for future studies on the effect of magnetic order and correlations on the vibrational spectra of materials with trirutile structure.

\section{Experimental and Calculation details}

The CuSb$_2$O$_6$ and CoSb$_2$O$_6$ crystals employed in this work were grown by chemical vapor transport as described elsewhere.\cite{Prokofiev,Rebello} The single crystals were oriented using Laue x-ray diffraction. Naturally grown {\it ab}- and {\it ac}- surfaces were employed in our Raman investigations on CuSb$_2$O$_6$ and CoSb$_2$O$_6$, respectively, with surface areas of the order of $1$ mm$^2$. The Raman spectra for CuSb$_2$O$_6$ and CoSb$_2$O$_6$ were taken in quasi-backscattering geometry using the 488.0 and 514.5 nm lines, respectively, of an Ar$^{+}$ laser as the exciting source, except where otherwise noted. The laser beam, with power below $\sim 10$ mW, was focused into a spot of $\sim 50$ $\mu$m diameter. The crystals were mounted on the cold finger of closed-cycle He cryostat with a base temperature of 20 K. A triple 1800 mm$^{-1}$ grating spectrometer equipped with a $L$N$_2$-cooled multichannel CCD detector was employed. The polarized data are represented in Porto's notation $A(BC)D$, where $B$ and $C$ indicate the electric field polarizations of the incident and scattered light, respectively, and $A$ and $D$ indicate the propagation directions of the incident and scattered light, respectively.

Our normal mode assignments and the corresponding mechanical representations were supported by $ab$-initio lattice dynamics calculations that were made using the PHonon package of the Quantum Espresso suite.\cite{QE} The calculations were carried out within the Density-Functional Perturbation Theory with the Perdew, Burke, and Enzerhof (PBE) Generalized Gradient Approximation (GGA) for the exchange-correlation potential\cite{Perdew} and the projetor augmented wave (PAW) method.\cite{Kresse,pseudopotentials} The energy cuttoffs for the wavefunctions and charge density were 40 and 200 Ry, respectively, and a $4 \times 4 \times 4$ grid for the reciprocal space was employed. The mechanical representations of the lattice vibrations were drawn using the program XCrySDen,\cite{xcrysden} while Fig. \ref{Structure} was prepared with the aid of the software VESTA.\cite{VESTA}

\section{Phonon Symmetry Analysis and First Principles Lattice Dynamics Calculations}

\begin{table}
	\caption{\label{ModesTet} Wyckoff positions and irreducible representations of $\Gamma$-point phonon modes for the high-symmetry tetragonal phase of trirutile structure, space group $P4_2/mnm$ ($D_{4h}^{14}$, No. 136). The corresponding Raman tensors are also given.}
	
	\begin{ruledtabular}
		\begin{tabular}{ccc}
			Atom & Wyckoff & $\Gamma$ -point phonon modes \\
			& position & \\
			\hline
			Cu & $2a$ & $A_{2u}$+$B_{1u}$+$2E_u$ \\
			Sb & $4e$ & $A_{1g}$+$A_{2u}$+$B_{1u}$+$B_{2g}$+$2E_g$+$2E_u$  \\
			O(1) & $4f$ & $A_{1g}$+$A_{2g}$+$A_{2u}$+$B_{1g}$+$B_{1u}$+$B_{2g}$+$E_g$+$2E_u$ \\
			O(2) & $8j$ & $2A_{1g}$+$A_{1u}$+$A_{2g}$+$2A_{2u}$+$B_{1g}$+$2B_{1u}$+$2B_{2g}$+ \\
			& & +$B_{2u}$+$3E_g$+$3E_u$ \\
			\hline
			\multicolumn{3}{c}{Classification:}  \\
			\multicolumn{3}{c}{$\Gamma_{Raman}$=$4A_{1g}$+$2B_{1g}$+$4B_{2g}$+$6E_g$} \\
			\multicolumn{3}{c}{$\Gamma_{IR}$=$5A_{2u}$+$9E_u$} \\
			\multicolumn{3}{c} {$\Gamma_{Silent}$=$A_{1u}$+$2A_{2g}$+$5B_{1u}$+$B_{2u}$} \\
			\multicolumn{3}{c} {$\Gamma_{Acoustic}$=$A_{2u}$+$E_u$} \\
			\hline
			\multicolumn{3}{c}{Raman tensors:} \\
			\multicolumn{3}{c}{ $A_{1g} \rightarrow
				\begin{pmatrix}
				a & 0 & 0 \\
				0 & a & 0 \\
				0 & 0 & b \\ 
				\end{pmatrix}$,
				$B_{1g} \rightarrow
				\begin{pmatrix}
				c & 0 & 0 \\
				0 & -c & 0 \\
				0 & 0 & 0 \\ 
				\end{pmatrix},
				B_{2g} \rightarrow
				\begin{pmatrix}
				0 & d & 0 \\
				d & 0 & 0 \\
				0 & 0 & 0 \\ 
				\end{pmatrix}$} \\
			\multicolumn{3}{c}{$E_{g} \rightarrow
				\begin{pmatrix}
				0 & 0 & 0 \\
				0 & 0 & e \\
				0 & e & 0 \\ 
				\end{pmatrix},
				\begin{pmatrix}
				0 & 0 & -e \\
				0 & 0 & 0 \\
				-e & 0 & 0 \\ 
				\end{pmatrix}$} \\
			
		\end{tabular}
	\end{ruledtabular}
\end{table}

\begin{table}
	\caption{\label{ModesMono} Similar to Table \ref{ModesTet}, for the low-symmetry monoclinic phase of the trirutile structure, space group $P2_1/n$ ($C_{2h}^{5}$, No. 14, unique axis $b$).}
	
	\begin{ruledtabular}
		\begin{tabular}{ccc}
			Atom & Wyckoff & $\Gamma$ -point phonon modes \\
			& position & \\
			\hline
			Cu & $2a$ & $3A_{u}$+$3B_{u}$ \\
			Sb & $4e$ & $3A_{g}$+$3A_{u}$+$3B_{g}$+$3B_{u}$  \\
			O(1) & $4e$ & $3A_{g}$+$3A_{u}$+$3B_{g}$+$3B_{u}$  \\
			O(2) & $4e$ & $3A_{g}$+$3A_{u}$+$3B_{g}$+$3B_{u}$  \\
			O(3) & $4e$ & $3A_{g}$+$3A_{u}$+$3B_{g}$+$3B_{u}$  \\
			\hline
			\multicolumn{3}{c}{Classification:}  \\
			\multicolumn{3}{c}{$\Gamma_{Raman}$=$12A_{g}$+$12B_{g}$} \\
			\multicolumn{3}{c}{$\Gamma_{IR}$=$15A_{u}$+$15B_u$} \\
			\multicolumn{3}{c}{$\Gamma_{Acoustic}$=$A_{u}$+$2B_u$} \\
			\hline
			\multicolumn{3}{c}{Raman tensors:} \\
			\multicolumn{3}{c}{$A_{g} \rightarrow
				\begin{pmatrix}
				a & 0 & d \\
				0 & b & 0 \\
				d & 0 & c \\ 
				\end{pmatrix}$,
				$B_{g} \rightarrow
				\begin{pmatrix}
				0 & e & 0 \\
				e & 0 & f \\
				0 & f & 0 \\ 
				\end{pmatrix}$} \\
			
		\end{tabular}
	\end{ruledtabular}
\end{table}

\begin{table}
	\caption{\label{Correlation} Correlation between the modes of the high-$T$ tetragonal phase (space group $D_{4h}^{14}$) and the low-$T$ monoclinic one (space group $C_{2h}^5$). The mode activities are also given.}
	
	\begin{ruledtabular}
		\begin{tabular}{ccccc}
			\multicolumn{2}{c}{Tetragonal} & & \multicolumn{2}{c}{Monoclinic} \\
			Symmetry & Activity & &  Symmetry & Activity \\
			\hline
			$A_{1g}$ & Raman & $\leftrightarrow$ & $A_g$ & Raman \\
			$A_{1u}$ & Silent & $\leftrightarrow$ & $A_u$ & IR \\
			$A_{2g}$ & Silent & $\leftrightarrow$ & $B_g$ & Raman \\
			$A_{2u}$ & IR & $\leftrightarrow$ & $B_u$ & IR \\
			$B_{1g}$ & Raman & $\leftrightarrow$ & $A_g$ & Raman \\
			$B_{1u}$ & Silent & $\leftrightarrow$ & $A_u$ & IR \\
			$B_{2g}$ & Raman & $\leftrightarrow$ & $B_g$ & Raman \\
			$B_{2u}$ & Silent & $\leftrightarrow$ & $B_u$ & IR \\
			$E_g$ & Raman & $\leftrightarrow$ & $A_g$+$B_g$ & Raman \\
			$E_u$ & IR & $\leftrightarrow$ & $A_u$+$B_u$ & IR \\
			
		\end{tabular}
	\end{ruledtabular}
\end{table}

\begin{table*}
	\caption{\label{Assignment} Observed frequencies of the Raman-active phonon modes of CuSb$_2$O$_6$ at $T=30$ K (monoclinic space group $P2_1/n$) and $T=450$ K (tetragonal space group $P4_2/mnm$) and tetragonal CoSb$_2$O$_6$ at $T=300$ K. The symmetry assignments for the modes in both space groups and calculated eigenfrequencies for tetragonal CuSb$_2$O$_6$ using either experimental non-relaxed \cite{Giere} or relaxed structural parameters as inputs (see text) are also given. Except for the silent modes at the tetragonal phase indicated by an (S), all listed phonons are Raman-active. The corresponding mechanical representations are drawn in Figs. \ref{Modes1} and \ref{Modes2}.}
	
	\begin{ruledtabular}
		\begin{tabular}{ccccccc}
			& & CuSb$_2$O$_6$ & CuSb$_2$O$_6$ & CuSb$_2$O$_6$ & CuSb$_2$O$_6$ & CoSb$_2$O$_6$ \\
			Assignment & Assignment & Calcul. (cm$^{-1}$) & Calcul. (cm$^{-1}$) & Observ. (cm$^{-1}$) & Observ. (cm$^{-1}$) & Observ. (cm$^{-1}$) \\
			& & non-relaxed & relaxed & $T=450$ K & $T=20$ K & $T=300$ K \\
			$P4_2/mnm$ & $P2_1/n$ & $P4_2/mnm$ & $P4_2/mnm$ & $P4_2/mnm$ & $P2_1/n$ & $P4_2/mnm$ \\
			\hline
			
			$B_{1g}(1)$ & $A_g(1)$ & imaginary & 110 & 73 & 98 & \\
			$E_{g}(1)$ & $A_g(2)+B_g(1)$ & 84 & 86 & & 113 & \\
			$E_{g}(2)$ & $A_g(3)+B_g(2)$ & 171 & 188 & & 226 &\\
			$A_{2g}(1)$ (S) & $B_g(3)$ & 216 & 228 & & & \\
			$B_{2g}(1)$ & $B_{g}(4)$ & 246 & 218 & 236 & 240 & \\
			$E_{g}(3)$ & $A_g(4)+B_g(5)$ & 260 & 241 & & 284, 286 & 277 \\
			$B_{1g}(2)$ & $A_g(5)$ & 284 & 290 & 313 & 318 & \\
			$A_{1g}(1)$ & $A_g(6)$ & 300 & 271 & 309 & 312 & 310\\
			$A_{2g}(2)$ (S) & $B_g(6)$ & 357 & 340 & & & \\
			$E_{g}(4)$ & $A_g(7)+B_g(7)$ & 413 & 374 & & 432 & 445 \\
			$A_{1g}(2)$ & $A_g(8)$ & 503 & 456 & 507 & 501 & 521 \\
			$B_{2g}(2)$ & $B_g(9)$ & 506 & 470 & 585 & 585 &\\
			$E_{g}(5)$ & $A_g(9)+B_g(8)$ & 605 & 498 & & 557 & 552 \\
			$A_{1g}(3)$ & $A_g(10)$ & 671 & 567 & 644 & 627 & 656 \\
			$B_{2g}(3)$ & $B_g(11)$ & 688 & 588 & & 700 & \\
			$E_{g}(6)$ & $A_g(11)+B_g(10)$ & 723 & 585 & & $676+653$ & \\
			$A_{1g}(4)$ & $A_g(12)$ & 744 & 624 & 721 & 730 & 725 \\
			$B_{2g}(4)$ & $B_g(12)$ & 843 & 703 & 816 & 831 & \\
			
		\end{tabular}
	\end{ruledtabular}
\end{table*}

\begin{table*}
	\caption{\label{Assignment2} Observed frequencies of the infrared-active phonon modes of CuSb$_2$O$_6$ as reported in Ref. \onlinecite{Torgashev} at $T=5$ and 300 K. The symmetry assignment according to our {\it ab}-initio lattice dynamical calculations using either experimental non-relaxed \cite{Giere} or relaxed structural parameters are also given. The silent modes at the tetragonal phase are indicated by (S). The corresponding mechanical representations are drawn in Fig. \ref{Modes3}.}
		\begin{ruledtabular}
		\begin{tabular}{cccccc}
			 &  & CuSb$_2$O$_6$ & CuSb$_2$O$_6$ & CuSb$_2$O$_6$ & CuSb$_2$O$_6$ \\
			Assignment & Assignment & Calcul. (cm$^{-1}$) & Calcul. (cm$^{-1}$) & Observ. (cm$^{-1}$) & Observ. (cm$^{-1}$)\\
			& & non-relaxed & relaxed & $T=5$ K & $T=300$ K \\
		    $P4_2/mnm$ & $P2_1/n$ & $P4_2/mnm$ & $P4_2/mnm$ & $P2_1/n$ & $P2_1/n$ \\
		    & & & & Ref. \onlinecite{Torgashev} & Ref. \onlinecite{Torgashev} \\

			\hline
			
			$E_u(1)$ (acoustic) & $A_u(1)+B_u(1)$ & & & & \\
			$A_{2u}(1)$ (acoustic) & $B_u(2)$ & & & & \\
			$E_u(2)$ & $A_u(2)+B_u(3)$ & 110 & 57 & $132.59+133.65$ & $132.02+133.35$ \\
			$A_{1u}(1)$ (S) & $A_{u}(3)$ & 112 & 162 & 169.48 & 171.17\\
			$B_{1u}(1)$ (S) & $A_u(4)$ & 155 & 148 & 177.30 & 177.92 \\
			$E_u(3)$ & $A_u(5)+B_u(4)$ & 197 & 166 & $202.38+203.10$ & $199.49+200.90$ \\
			$E_u(4)$ & $A_u(6)+B_u(5)$ & 213 & 216 & $226.78+225.51$ & $225.51$ ($B_u$)\\
			$E_u(5)$ & $A_u(7)+B_u(6)$ & 254 & 242 & $264.04+259.70$ & $263.32+259.10$ \\
			$B_{1u}(2)$ (S) & $A_u(8)$ & 258 & 234 & 238.05 &  \\
			$A_{2u}(2)$ & $B_u(7)$ & 262 & 245 & 259.70 & 259.10 \\
			$E_u(6)$ & $A_u(9)+B_u(8)$ & 292 & 275 & $283.90+286.42$ & $ 280.76 + 285.52$ \\
			$B_{2u}(1)$ (S) & $B_u(9)$ & 334 & 313 & 325.67 & $324.17$ \\
			$A_{2u}(3)$ & $B_u(10)$ & 483 & 424 & 547.00 & 547.00 \\
			$E_u(7)$ & $A_u(10)+B_u(11)$ & 491 & 443 & $485.93+490.00$ & $486.53+490.00$ \\
			$B_{1u}(3)$ (S) & $A_u(11)$ & 505 & 456 & 315.24 & 315.29 \\
			$E_{u}(8)$ & $A_u(12)+B_u(12)$ & 593 & 525 & $566.16+563.00$ & $567.36+562.50$ \\
			$A_{2u}(4)$ & $B_u(13)$ & 650 & 550 & 591.50 & 591.50\\
			$B_{1u}(4)$ (S) & $A_u(13)$ & 679 & 563 & 653.94 & 654.64 \\
			$E_u(9)$ & $A_u(14)+B_u(14)$ & 730 & 597 & $737.65+735.50$ & $737.51+734.33$ \\
			$B_{1u}(5)$ (S) & $A_u(15)$ & 745 & 624 & 715.11 & 707.31 \\
			$A_{2u}(5)$ & $B_u(15)$ & 836 & 700 & 842.00 & 832.47 \\
			
		\end{tabular}
	\end{ruledtabular}
\end{table*}

\begin{figure*}
	\includegraphics[width=1.0 \textwidth]{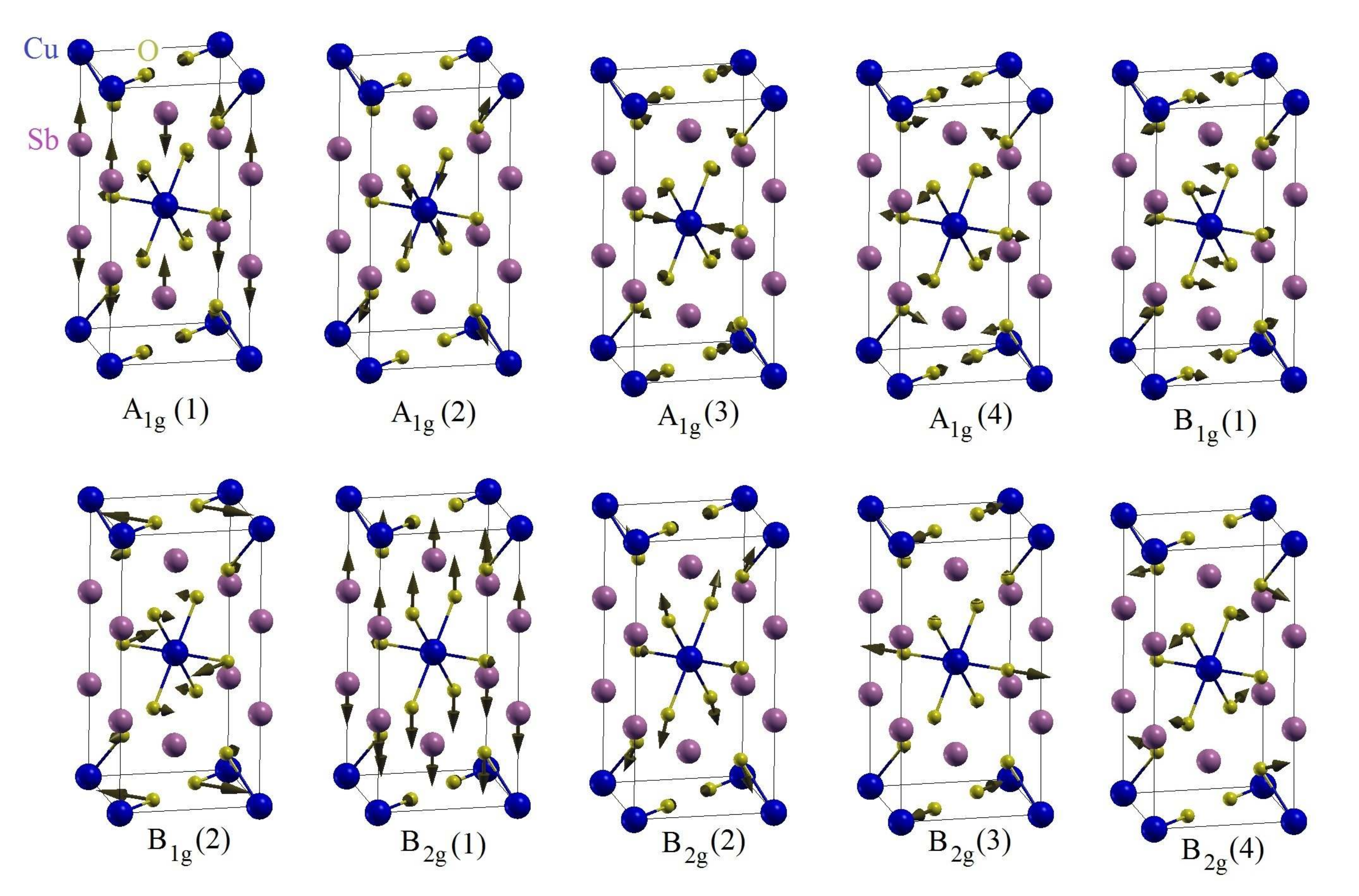}
	\caption{\label{Modes1} Mechanical representations of the Raman-active $A_{1g}$, $B_{1g}$, and $B_{2g}$ modes of the trirutile structure (tetragonal unit cell).}
\end{figure*}

\begin{figure*}
	\includegraphics[width=1.0 \textwidth]{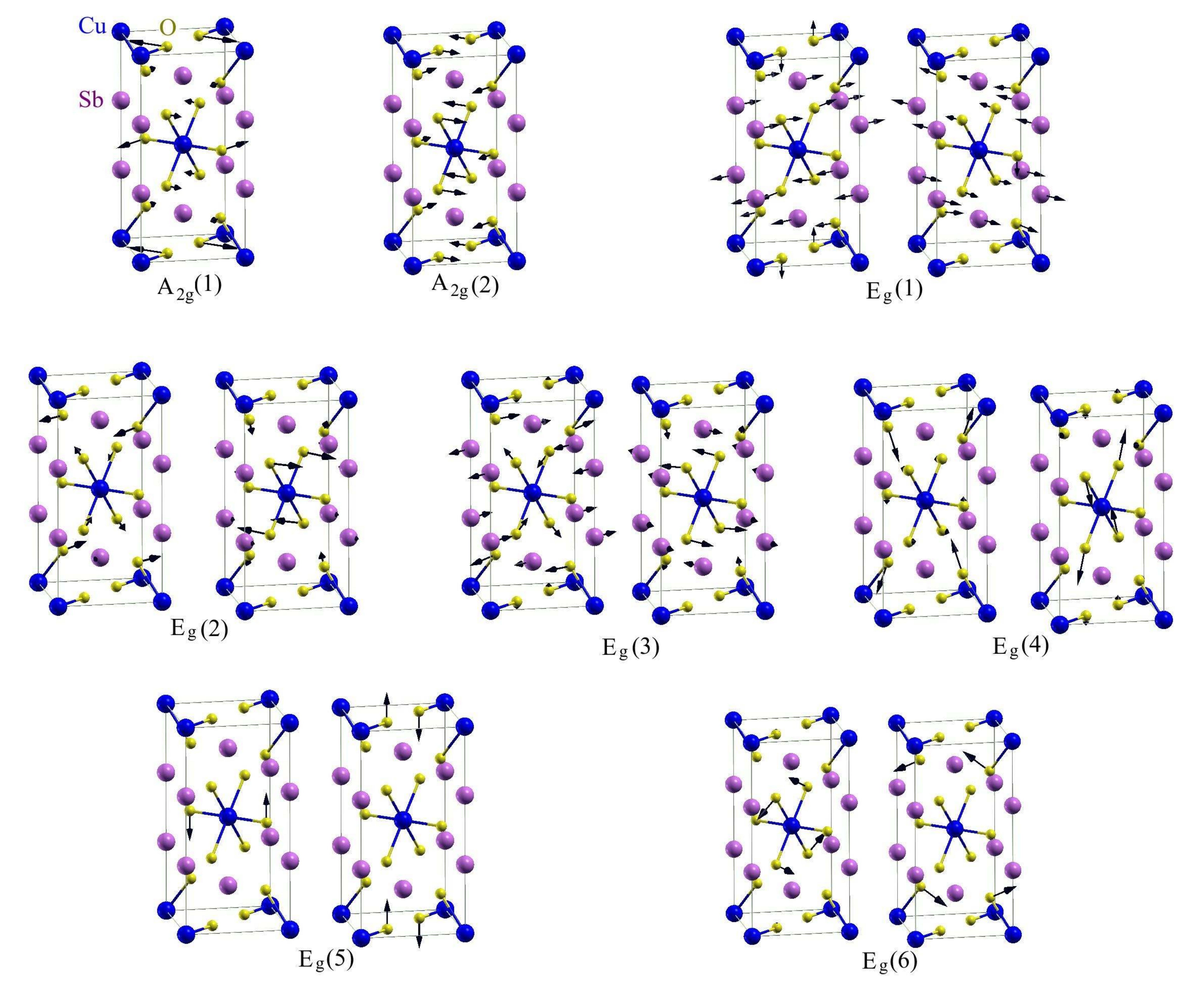}
	\caption{\label{Modes2} Mechanical representations of the silent $A_{2g}$ and Raman-active $E_{g}$ modes of the trirutile structure (tetragonal unit cell).}
\end{figure*}

\begin{figure*}
	\includegraphics[width=0.9 \textwidth]{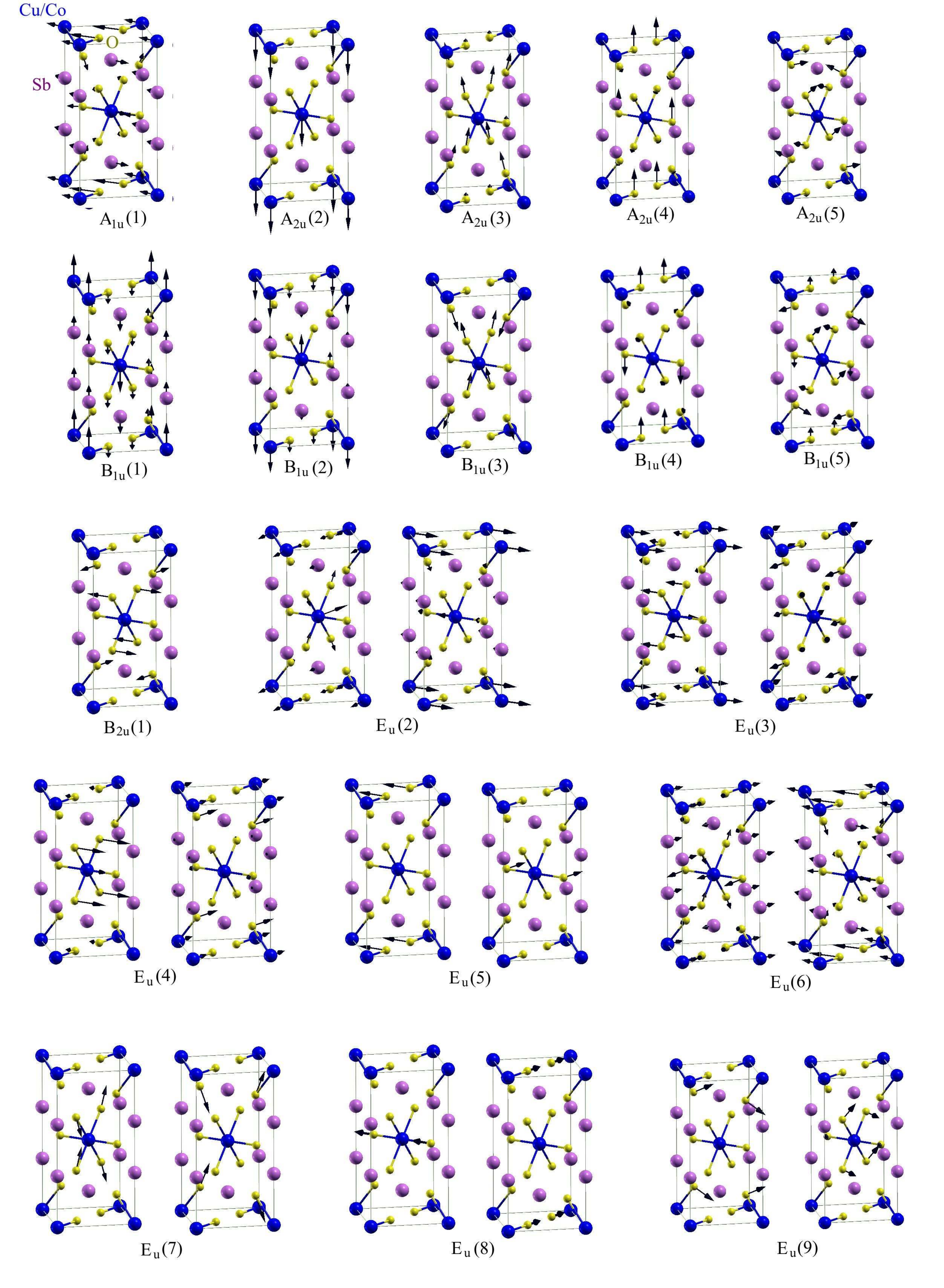}
	\caption{\label{Modes3} Mechanical representations of the odd ({\it ungerade}) optical modes of the trirutile structure (tetragonal unit cell).}
\end{figure*}

Both the tetragonal (space group $P4_2/mnm$) and monoclinic (space group $P2_1/n$) variants of the trirutile structure present two chemical formulas per unit cell. The irreducible representations of the $\Gamma$-point phonon modes as well as the corresponding Raman tensors for the tetragonal and monoclinic phases are listed in Tables \ref{ModesTet} and \ref{ModesMono}, respectively. The correlation between the modes in the tetragonal and monoclinic phases are shown in Table \ref{Correlation}. In total, 16 Raman-active modes are predicted for the tetragonal phase ($\Gamma_{Raman}^{tetragonal} =4A_{1g}$+$2B_{1g}$+$4B_{2g}$+$6E_g$), while 24 Raman modes are expected for the monoclinic structure ($\Gamma_{Raman}^{monoclinic} =12A_{g}$+$12B_{g}$). The higher number of Raman-active modes in the monoclinic phase arise from the splitting of the $E_g$ modes into $A_g$+$B_g$ modes and also from the fact that the two $A_{2g}$ silent modes of the tetragonal phase become Raman-active $B_g$ modes after the monoclinic distortion.

{\it Ab}-initio lattice dynamics calculations were performed for the tetragonal phase of CuSb$_2$O$_6$, following two independent procedures: (i) the experimental lattice parameters and atomic positions reported in Ref. \onlinecite{Giere} were employed as inputs for the calculation; and (ii) a relaxation of the atomic positions and lattice parameters was performed prior to the lattice dynamics calculation to reach the minimal total energy. Tables \ref{Assignment} and \ref{Assignment2} show the frequencies of the even ({\it gerade}) and odd ({\it ungerade}) modes, respectively, calculated using both procedures, and the corresponding symmetries for the tetragonal and monoclinic space groups. The mechanical representations of these modes are shown in Figs. \ref{Modes1}, \ref{Modes2}, and \ref{Modes3}. An assignment of the previously reported infrared-active modes at 5 and 300 K \cite{Torgashev} according to our lattice dynamics calculations is also given in Table \ref{Assignment2}.

\section{Experimental Results and Analysis}

\begin{figure}
	\includegraphics[width=0.5 \textwidth]{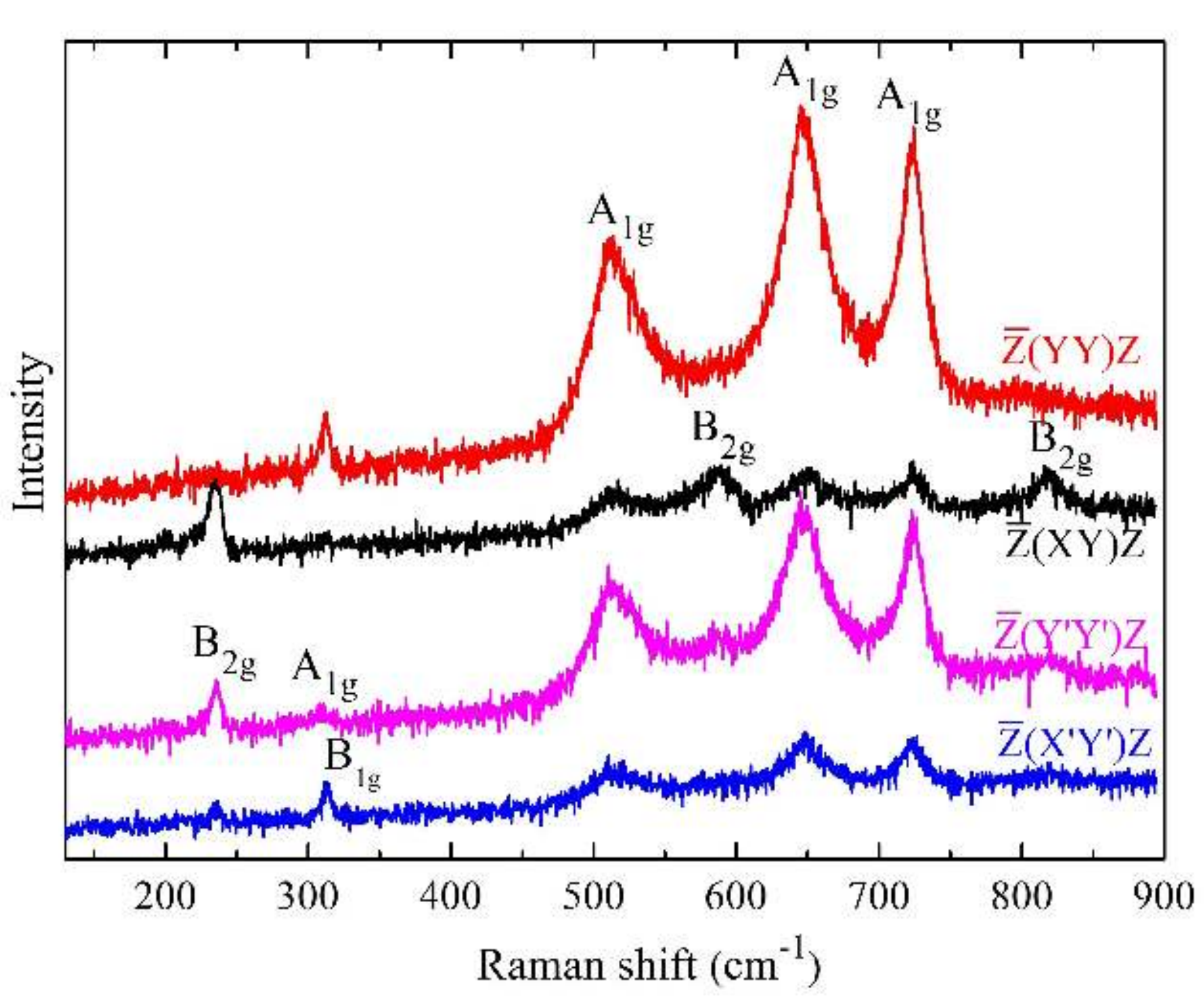}
		\caption{\label{pol} (a) Polarized Raman spectra of CuSb$_2$O$_6$ at $T=450$ K. X' and Y' represent diagonal directions in the {\it ab} plane. This experiment was performed with the $\lambda = 514.5$ nm Ar$^{+}$ laser line. Spectra were vertically translated for clarity.}
\end{figure}

\begin{figure}
	\includegraphics[width=0.5 \textwidth]{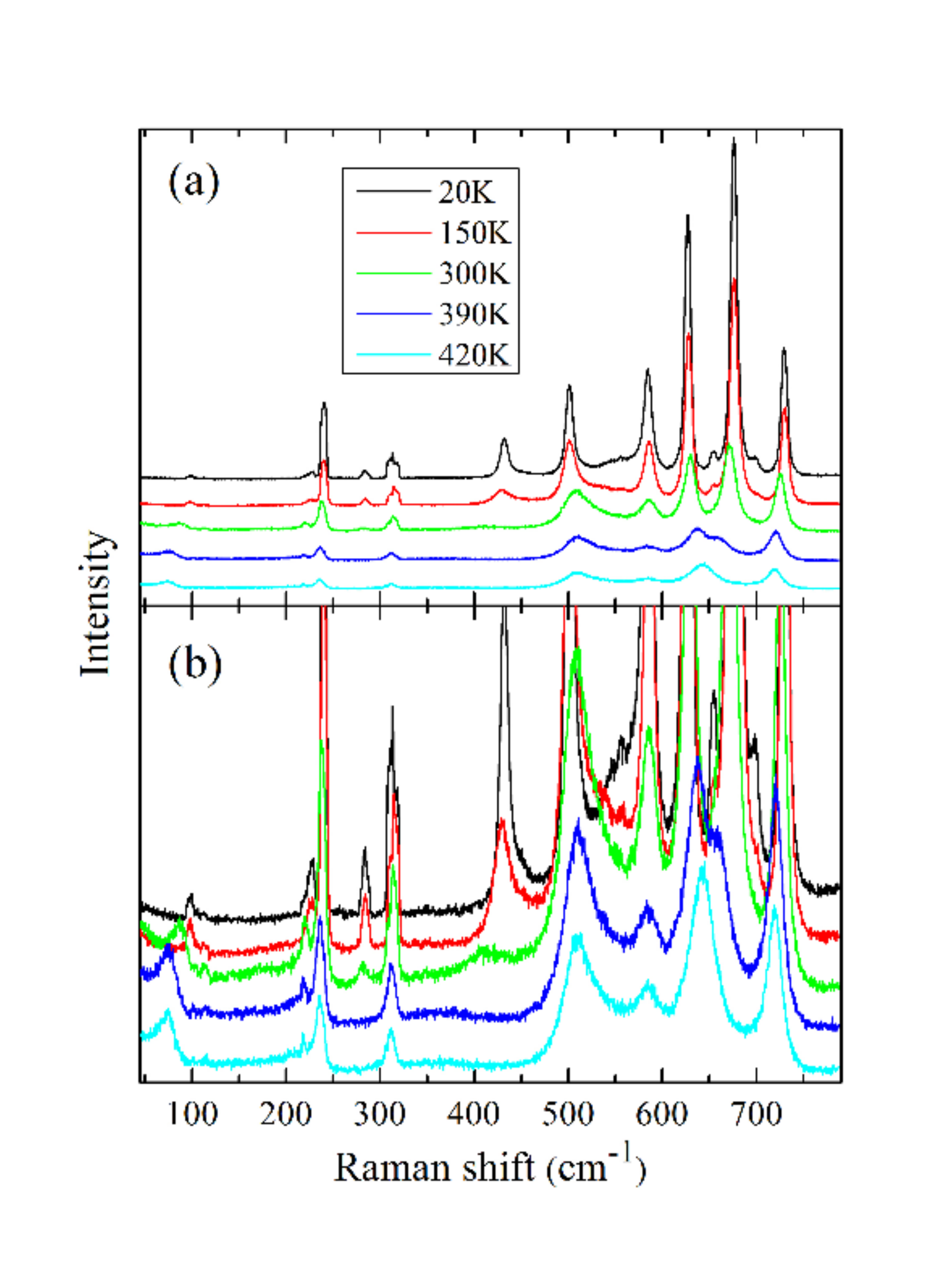}
		\caption{\label{Spectra_Tdep} (a) Unpolarized Raman spectra of CuSb$_2$O$_6$ at selected temperatures. (b) Same as (a), in an expanded intensity scale. Spectra were vertically translated for clarity. The weak sharp peak at 221 cm$^{-1}$ is a laser plasma line.}
\end{figure}

\begin{figure}
	\includegraphics[width=0.5 \textwidth]{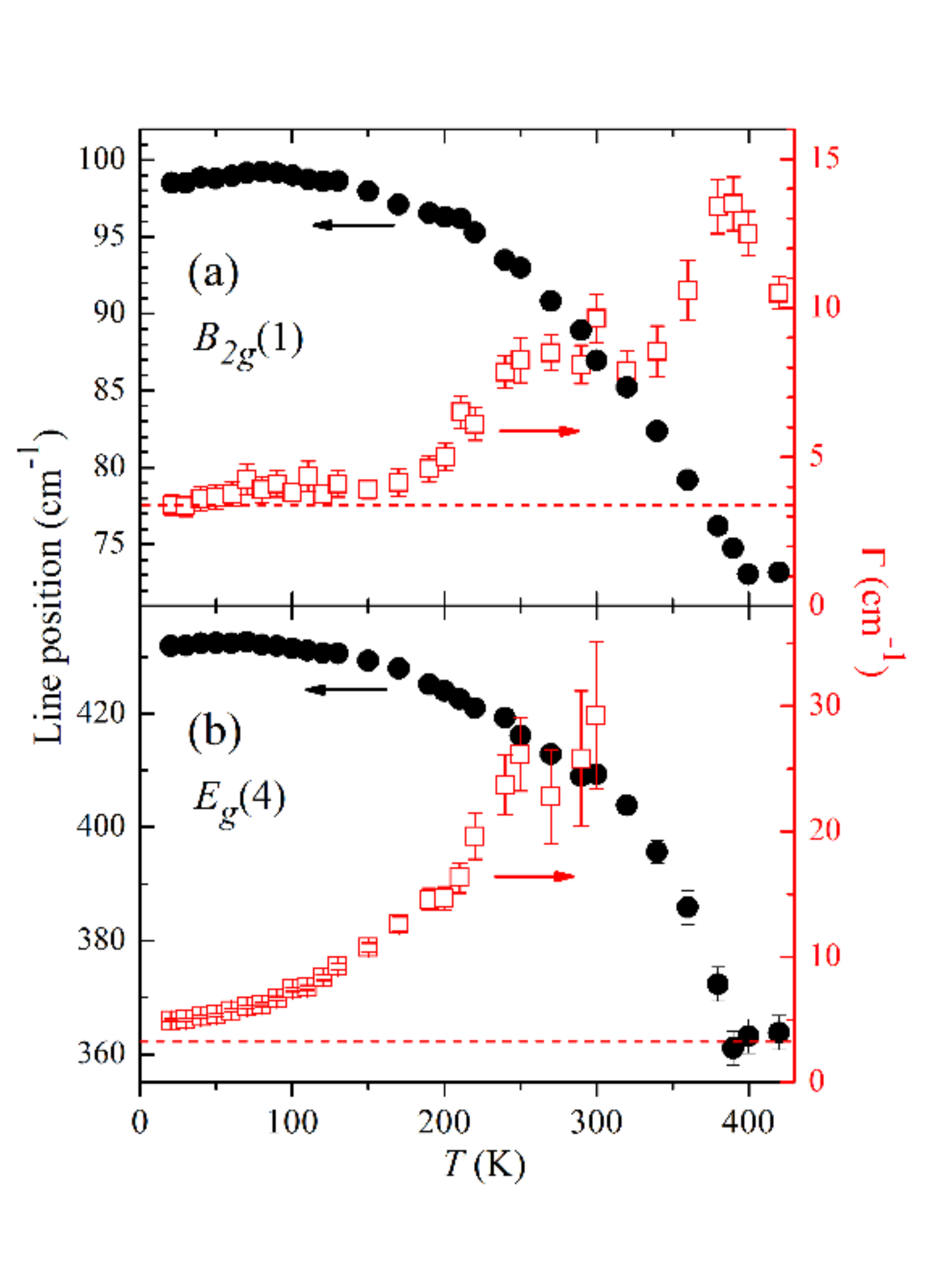}
	\caption{\label{soft} Temperature-dependence of the line positions and damping constant $\Gamma$ (corresponding to the half width at half maximum) of the $B_{1g}(1)$ (a) and $E_g(4)$ (b) modes of CuSb$_2$O$_6$. The instrumental linewidth for this experiment, $\Gamma_{inst}=3.4$ cm$^{-1}$, is indicated by horizontal dashed lines.}
\end{figure}

\begin{figure}
	\includegraphics[width=0.5 \textwidth]{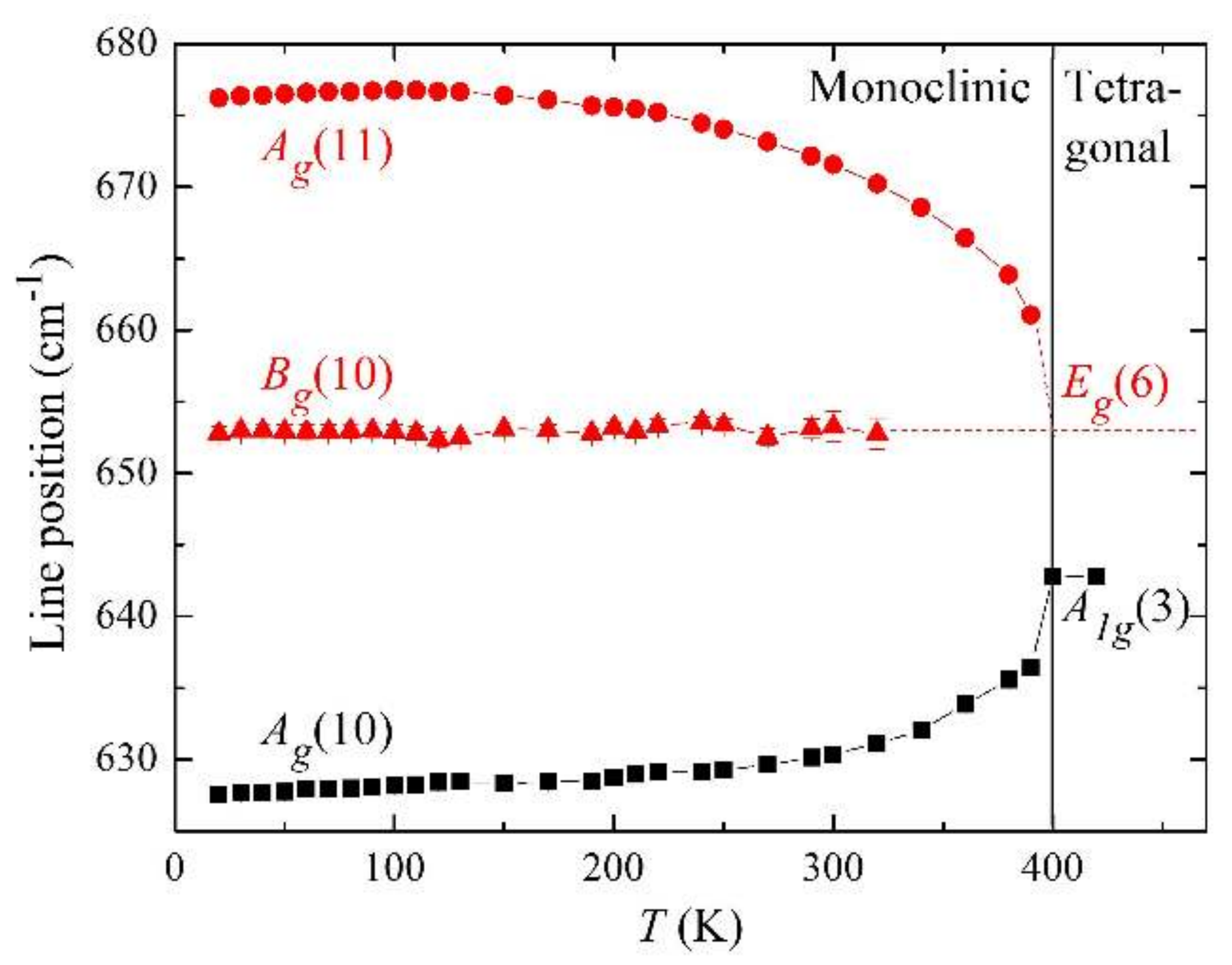}
	\caption{\label{Split} Temperature-dependence of the observed peak positions in the spectral region $620 < \omega < 680$ cm$^{-1}$ of CuSb$_2$O$_6$. The vertical solid line mark the structural transition temperature $T_s=400$ K.}
\end{figure}

\begin{figure}
	\includegraphics[width=0.5 \textwidth]{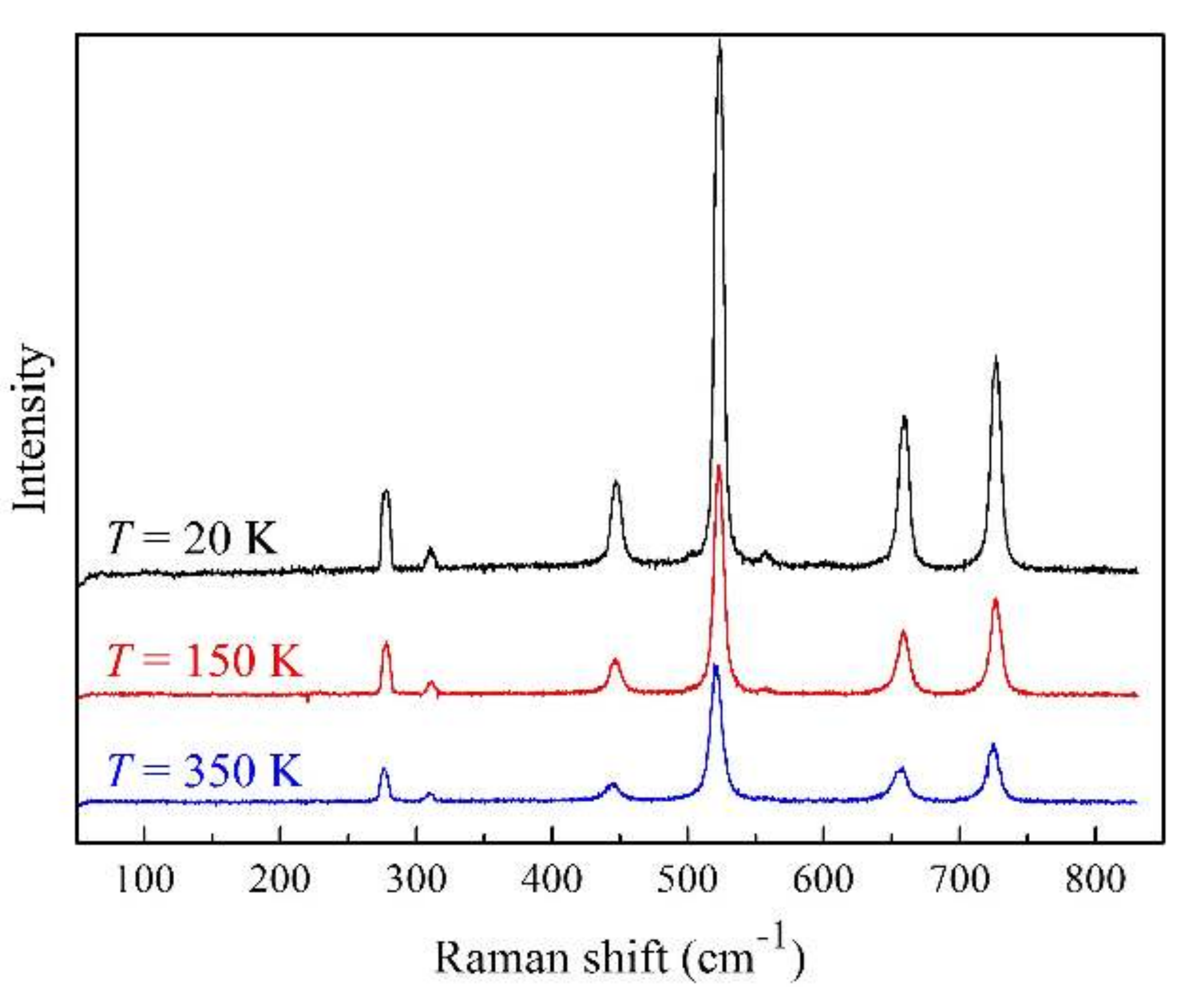}
	\caption{\label{CoSb2O6} Unpolarized Raman spectra of CoSb$_2$O$_6$ at selected temperatures. Spectra were vertically translated for clarity.}
\end{figure}

Figure \ref{pol} shows the linearly polarized Raman spectra of tetragonal CuSb$_2$O$_6$ at $T=450$ K. Three $A_{1g}$ modes, at 507, 644, and 721 cm$^{-1}$, are identified for their strong intensities in $\bar{Z}(YY)Z$ and $\bar{Z}(Y'Y')Z$ polarizations. In addition, a $B_{1g}$ mode is observed in $\bar{Z}(YY)Z$ and $\bar{Z}(X'Y')Z$ polarizations at 313 cm$^{-1}$ [$B_{1g}(2)$], next to a weaker $A_{1g}$ mode at 309 cm$^{-1}$. Finally, three $B_{2g}$ modes were observed at 236, 585, and 816 cm$^{-1}$. Residual intensities of the strong $A_{1g}$ modes were also observed in the forbidden $\bar{Z}(XY)Z$ and $\bar{Z}(X'Y')Z$ polarizations. This partial depolarization effect is ascribed to a slight deviation of our experimental geometry with respect to the ideal backscattering geometry ($\theta_{in} \sim 10^{\circ}$), allied with a possible birefringence of the sample.

Figures \ref{Spectra_Tdep}(a) and \ref{Spectra_Tdep}(b) show unpolarized Raman spectra of CuSb$_2$O$_6$ at selected temperatures. At low temperatures, additional Raman peaks are detected at 113, 226, 284, 286, 432, 557, 653, 676, and 700 cm$^{-1}$. The assignments of all observed modes at the tetragonal and monoclinic phases of CuSb$_2$O$_6$, supported by a comparison to the results of our {\it ab}-initio lattice dynamics calculations, are given in Table \ref{Assignment}. The extra modes at 113, 226, 284, 286, 432, 557, 653 and 676 cm$^{-1}$ are monoclinic $A_g$ and/or $B_g$ modes derived from tetragonal $E_g$ modes that are not observed in the {\it ab}-plane quasi-backscattering geometry of this experiment in the high-symmetry phase, and the 700 cm$^{-1}$ peak is ascribed to a $B_g$ mode derived from a weak tetragonal $B_{2g}$ mode that is not observed at high temperatures within our sensitivity. Finally, the mode at 73 cm$^{-1}$ at $T=420$ K (98 cm$^{-1}$ at $T=20$ K) is ascribed to a $B_{1g}$ ($A_g$) mode in the tetragonal (monoclinic) phase.

Figures \ref{soft}(a) and \ref{soft}(b) show the temperature-dependence of the peak positions of the $B_{1g}(1)$ and $E_g(4)$ modes (tetragonal cell), associated with rotations around the $c$ axis and Jahn-Teller-like elongations, respectively, of the CuO$_6$ octahedra (see also Figs. \ref{Modes1} and \ref{Modes2}). The structural transition is clearly manifested by remarkable softenings of 26\% and 17\%, respectively, of these modes on warming up to $T_s=400$ K, indicating these would be the soft modes of the transition. Nonetheless, we emphasize that these modes do not freeze ($\omega \rightarrow 0$) as $T \rightarrow T_s$ as would be expected for a classical soft mode behavior.\cite{Cochran} The linewidth $\Gamma(T)$ of the $B_{1g}(1)$ mode is resolution-limited at low temperatures ($\Gamma_{inst}=3.4$ cm$^{-1})$, broadens to $\Gamma \sim 8$ cm$^{-1}$ between 250 and 350 K, and reaches its maximum value $\Gamma \sim 14$ cm$^{-1}$ near $T_s$ [see Fig. \ref{soft}(a)]. We should mention that, although the $E_g(4)$ mode is forbidden for {\it ab}-plane scattering, a residual intensity is still observed in the tetragonal phase [see Fig. \ref{Spectra_Tdep}(b)], which was sufficient to determine its peak position in the whole investigated temperature interval. Nonetheless, the reliable determination of $\Gamma$ for this mode could only be obtained up to $\sim 300$ K, showing a qualitatively similar behavior with respect to the $B_{1g}(1)$ mode [see Fig. \ref{soft}(b)].

The structural phase transition at $T_s$ is also manifested by large spectral changes at the spectral region $\omega \sim 620-680$ cm$^{-1}$, consistent with a previous report.\cite{Giere} Above $T_s$, a single mode is observed at 644 cm$^{-1}$, while two distinct modes are observed immediately below $T_s$ [see Fig. \ref{Spectra_Tdep}(a)]. Also, a third and much weaker mode at 653 cm$^{-1}$ is detected at low temperatures. Figure \ref{Split} shows the temperature-dependence of the position of these peaks. As $T \rightarrow T_s$ from below, the two strongest modes approach each other. The separation between these modes does not tend to zero but to $\Delta \omega = 10(2)$ cm$^{-1}$ at $T_s=400$ K. We conclude that the lower-frequency mode in the monoclinic phase is the $A_g(10)$ mode originated from the 644 cm$^{-1}$ tetragonal $A_{1g}(3)$ mode, while the intermediate- and high-frequency peaks originate from a distinct [$E_g(6)$] mode at $\sim 653$ cm$^{-1}$ that is Raman-forbidden at $ab$-plane backscattering geometry in the tetragonal phase (see Table \ref{Assignment}). Below $T_s$, the tetragonal $E_g(6)$ mode transform in the $A_g(11)+B_g(10)$ modes. The so-derived $A_g(11)$ mode would then have the same symmetry of the $A_g(10)$ mode originated from the $A_{1g}(3)$ phonon in the tetragonal phase. The similar frequencies of these two $A_g$ phonons would favor their mixing \cite{Yan} in the monoclinic phase, leading to the observed ``repulsion'' of these modes on cooling below $T_s$. The weak peak at 653 cm$^{-1}$ is attributed to the $B_g(10)$ mode, also derived from the tetragonal $E_g(6)$ mode, but which does not mix with the $A_g(10)$ mode for having a different symmetry.

It is interesting to note that, despite the subtle monoclinic distortion of CuSb$_2$O$_6$ below $T_s$,\cite{Giere} the observed symmetry-breaking-induced monoclinic $A_g(10)+A_g(11)$ phonon mixing in CuSb$_2$O$_6$ is an extraordinarily large effect. We should mention that this large separation of the $\sim 650$ cm$^{-1}$ band in the monoclinic phase was also reported in Ref. \onlinecite{Giere} but rather interpreted in terms of a splitting of a degenerate $E_g$ mode. This interpretation is dismissed by our polarized Raman data at 450 K, which shows unambiguously that the 644 cm$^{-1}$ mode has $A_{1g}$ symmetry in the tetragonal phase (see Fig. \ref{pol}). 

The Raman spectra of tetragonal CoSb$_2$O$_6$ at selected temperatures are given in Fig. \ref{CoSb2O6}. Peaks at 277, 310, 445, 521, 552, 656, and 725 cm$^{-1}$ are observed, consistent with a previously published powder Raman spectrum at room temperature.\cite{Haeuseler} The mode assignment is given Table \ref{Assignment}. In contrast to the strong temperature-dependence observed for the Raman spectrum of CuSb$_2$O$_6$, little spectral changes were observed for CoSb$_2$O$_6$ between $20$ and $350$ K.

\section{Discussion}

\subsection{Lattice dynamical calculations and mode assignment}

Except for some modes below $\sim 300$ cm$^ {-1}$, the calculated phonon frequencies using the relaxed crystal structure of CuSb$_2$O$_6$ yielded softer modes as compared to the calculations using the experimental crystal structure (see Tables \ref{Assignment} and \ref{Assignment2}). The difference becomes more significant for the hardest modes, where the comparison with the observed frequencies favors the calculation using the experimental crystal structure. On the other hand, such calculation yielded a imaginary frequency for the $B_{1g}(1)$ soft mode (see Table \ref{Assignment}). Still, the comparison between experimental and calculated phonon frequencies is fair enough to allow for an assignment of the observed modes for CuSb$_2$O$_6$ with very little ambiguity. Also, the symmetry classification of observed modes in the tetragonal phase using polarized data (see Fig. \ref{pol}) resolved most of the ambiguities related with the existence of distinct modes at similar frequencies. The symmetry classification of the $A_{1g}(3)$ mode at 644 cm$^{-1}$ in the tetragonal phase (rather than an $E_g$ mode as previously assumed)\cite{Giere} was particularly important to interpret the apparent splitting of this mode in the monoclinic phase. 

Since {\it ab}-initio lattice dynamics calculations such as those carried out in this work do not deal with manual adjustments of force constants, less ambiguity in the mode assignment with respect to other calculation methods \cite{Husson,Haeuseler} arises for a relatively complex structure such as the trirutiles. Also, mode assignments through correlations between parent rutile and the trirutile phonons such as performed in Ref. \onlinecite{Torgashev} do not take into account the effect of distinct chemical species at the Cu and Sb sites on the normal modes, and may lead to largely different mechanical representations of the resulting vibrations for many phonons, as verified by a comparison between the drawings of Figs. \ref{Modes1}, \ref{Modes2} and \ref{Modes3} with those given in Ref. \onlinecite{Torgashev}. Overall, we believe that our methodology will be useful to interpret the vibrational spectra of other similar materials.

\subsection{Structural transition of CuSb$_2$O$_6$}

The Raman spectrum of CuSb$_2$O$_6$ shows a rich temperature dependence (see Fig. \ref{Spectra_Tdep}), in contrast to CoSb$_2$O$_6$ (see Fig. \ref{CoSb2O6}). Some observed effects for CuSb$_2$O$_6$ are clearly related with the Jahn-Teller structural transition at $T_s=400$ K. First of all, the phonons derived from the tetragonal $E_g$ modes fade away or show only residual intensities above $T_s$ in our Raman measurements, including the $E_g(4)$ soft mode. This is due to the specific geometry of our experiment, which exploited a natural {\it ab}-surface, with null $E_g$ Raman tensor elements in the tetragonal phase (see Table \ref{ModesTet}). Another interesting effect associated with the transition at $T_s$ is the remarkable phonon mixing at $\sim 650$ cm$^{-1}$ in the monoclinic phase, as described in detail in Section IV.

A revealing manifestation of the structural transition of CuSb$_2$O$_6$ in our data is the observation of large softenings of the peaks derived from the $B_{1g}(1)$ and $E_g(4)$ tetragonal modes on warming up to $T_s$ (see Fig. \ref{soft}). In fact, the $B_{1g}(1)$ mode softens from 99 cm$^{-1}$ at 20 K to 73 cm$^{-1}$ at 420 K (26\%), while the central position of the peak derived from the $E_g(4)$ tetragonal mode softens from 432 cm$^{-1}$ at 20 K to 360 cm$^{-1}$ at 420 K (17\%). These modes correspond, respectively, to a dynamic rotation of the CuO$_6$ octahedra along the $c$ axis and to Jahn-Teller-like elongations of the same octahedra. Considering that the structural differences between the $\alpha$-CuSb$_2$O$_6$ and $\beta$-CuSb$_2$O$_6$ phases are precisely a Jahn-Teller distortion plus a rotation of the CuO$_6$ octahedra (see Fig. \ref{Structure}), the $B_{1g}(1)$ and $E_g(4)$ modes could be indeed anticipated as good candidates for soft modes of the structural transition. On the other hand, the behavior of these modes is not characteristic of classical soft modes, which should freeze ($\omega \rightarrow 0$) as $T \rightarrow T_s$ (Ref. \onlinecite{Cochran}). A large anomalous broadening was observed for the $B_{1g}(1)$ mode on warming up to $T_s$, indicating an important influence of anharmonicity in this phase transition. In fact, the damping constant of the $B_{1g}(1)$ mode, which is resolution-limited at low temperatures, reach a maximum value of 14(1) cm$^{-1}$ at $T \sim T_s$ [see Fig. \ref{soft}(a)], corresponding to a phonon lifetime of only $\sim 3$ times the vibrational period. A large broadening is also observed for the $E_g(4)$ mode [see Fig. \ref{soft}(b)]. Such large phonon damping signals a crossover from a displacive to an order-disorder phase transition,\cite{Bruce,Muller} where incoherent local Jahn-Teller distortions and rotations of CuO$_6$ octahedra grow and dominate the lattice dynamics as $T \rightarrow T_s$. We conclude that, although the Jahn-Teller energy level splitting is slightly larger than the stiffness parameter that orients the distortion cooperatively, these quantities show comparable magnitudes. Also, the extrapolated behavior of the line positions of the ``soft'' modes indicate that, in the absence of the order-disorder transition at $T_s$, these modes would actually freeze ($\omega \rightarrow 0$) at $T \sim 600$ K. Based on these considerations, we suggest that local Jahn-Teller distortions of CuO$_6$ octahedra take place between $T_s$ and $\sim 600$ K, being likely washed out above this temperature. Further investigations at higher temperatures than those reached in this work are necessary to confirm or dismiss this scenario.

\section{Conclusions}

In summary, a Raman scattering study of CuSb$_2$O$_6$ and CoSb$_2$O$_6$ crystals with trirutile structure is presented, supported by {\it ab}-initio lattice dynamics calculations. The methodology employed in this work allowed for the identification of the observed Raman-active and previously published infrared-active modes \cite{Torgashev} of CuSb$_2$O$_6$, which proved to be determinant to interpret the observed phonon anomalies associated with the Jahn-Teller structural transition at $T_s=400$ K. A crossover from a displacive to an order-disorder transition at $T_s$ is inferred from our data, indicating that local Jahn-Teller fluctuations should be present immediately above $T_s$, with a possible modification to a state with regular local CuO$_6$ octahedra above $\sim 600$ K. A study of the effect of magnetic correlations on the vibrational Raman spectra of these materials is currently underway.\cite{Damaris}

\begin{acknowledgments}

Work at UNICAMP was conducted with financial support from FAPESP Grant 2012/04870-7, CAPES, and CNPq, Brazil. Work at Montana State University was conducted with financial support from the US Department of Energy (DOE), Office of Science, Basic Energy Sciences (BES) under Award No. DE-SC0016156.

\end{acknowledgments}

\end{document}